\documentclass[british,english]{article}
\usepackage[T1]{fontenc}
\usepackage[latin9]{inputenc}
\usepackage{geometry}
\usepackage{array}
\usepackage{float}
\usepackage{multirow}
\usepackage{esint}

\makeatletter

\providecommand{\tabularnewline}{\\}

\makeatother

\usepackage{babel}
\begin{document}
\title{Adjusting for publication bias in meta-analysis with continuous outcomes:
a comparative study}
\author{Osama Almalik\thanks{Researcher, Department of Mathematics and Computer Science, Eindhoven
University of Technology. E-mail: O.Almalik@tue.nl.}, }
\maketitle
\begin{abstract}
\noindent Publication bias has been a problem facing meta-analysts.
Methods adjusting for publication bias have been proposed in the literature.
Comparative studies for methods adjusting for publication bias are
found in the literature, but these studies are limited. We investigated
and compared the performance of five methods adjusting for publication
bias for the case of continuous outcomes. The methods studied are
Copas, PET-PEESE, p-uniform, Trim \& Fill and the limit meta-analysis.
In addition, the performance of the random-effects meta-analysis using
the DerSimonian estimator is also investigated. The analysis was done
using a case-study and an extensive simulation study including different
scenario's. The Copas and the PET-PEESE were found to be the least
biased methods adjusting for publication bias. However, the Copas
method, like other Likelihood-based methods, can have convergence
issues. In addition, the PET-PEESE method is robust in case of heteroscedasticity,
making the PET-PEESE method a preferable technique to adjust for publication
bias.
\end{abstract}

\section{Introduction}

\selectlanguage{british}%
\noindent Publication bias, also known as the file drawer problem
or small study problem, occurs when the published meta-analysis is
not representative of the population of all studies, and can lead
to incorrect conclusions. Publication bias is mainly attributed to
the direction and the statistical significance of results, but other
causes have been documented in the literature, e.g. language bias,
availability bias and cost bias \cite{key-1-1-1-1-1-1}. 

\selectlanguage{english}%
\noindent One popular method for adjustment for publication bias is
the Trim and Fill method \cite{key-1-1-1-1-1-2,key-1-1-1-1-1-3}.
Assuming funnel plot symmetry, the method starts by estimating the
number of unreported studies. Then the treatment effect is re-estimated
after removing the most extreme studies on the left side of the funnel
plot. Using the new treatment effect estimate, a new number of unreported
studies is estimated and removed. This process is repeated till the
number of unreported studies is stable. Then values are imputed for
this number of studies on the right side of the funnel plot, and a
treatment effect is estimated using the new set of data. The method
had been evaluated in the literature for binary outcomes, and was
found to be more biased in case of treatment effect heterogeneity
\cite{key-1-1-1-1-1-4,key-1-1-1-1-1-5}. 

\selectlanguage{british}%
\noindent Methods correcting for publication bias applying a selection
model assume a mechanism for the publication process while modelling
how the data are generated without publication bias. The first selection
method developed by Hedges (1984) assumes that only statistically
significant results are published \cite{key-1-1-1-1-1-1}. Iyengar
and Greenhouse (1988) developed the Hedges method by using a weight
function based on the relative likelihood that non-significant result
study is published to that of a significant result study \cite{key-1-1-1-1-1}.
Later on the Hedges and Iyengar and Greenhouse methods have been extended
to accommodate more complicated forms of publication bias \cite{key-1-1-1-1-2}.
Vevea and Hedges proposed creating weights using a step function involving
the one-tailed p-values \cite{key-1-1-1-1-2}. The Vevea and Hedges
method is computationally complex since all weights need to be estimated
along with the effect size and the between-study variance component.
This can be especially problematic when the number of studies in a
meta-analysis is small. For this reason, Vevea and Woods (2005) proposed
applying the Vevea and Hedges method with apriori specified weight
functions. However, one disadvantage of the Vevea and Hedges method
is assuming the study-specific covariates are available \cite{key-1-1-1-1-3-1}.
\foreignlanguage{english}{Dear and Begg (1992) introduced a similar
method using a weighted Likelihood function with the weights calculated
using limits from the normal distribution based on each study's p
value \cite{key-1-1-1-1-3-2}.}

\noindent Another adjustment method based on a selection model which
uses the effect size and its standard error instead of the p values
was developed as a sensitivity analysis by Copas and Shi \cite{key-1-1-1-1-4,key-1-1-1-1-5}.
The probability of selection is estimated using a linear function
of a standard normally distributed variable, with the function parameters
determined subjectively. Next, a conditional distribution density
function is constructed based on the distribution of the treatment
effects and of the selection probability, after which Likelihood methods
are applied to obtain estimates for the treatment effect. The sensitivity
of the estimated treatment effect values is tested for different values
of the parameters of the selection function. Reasonable estimates
for the treatment effect are the ones associated with a p-value of
a publication bias test of just above 0.1.\foreignlanguage{english}{ }

\selectlanguage{english}%
\noindent Another group of selection methods uses the distribution
of the p-values \cite{key-1-1-1-1-6,key-1-1-1-1-7}. The p-curve and
p-uniform methods both estimate the effect size by minimizing the
distance between the observed p-values distribution and the uniform
distribution, each method using a different distance metrics. One
disadvantage of the p curve method is that it does not estimate a
confidence interval for the bias-adjusted effect size \cite{key-1-1-1-1-6}.
More selection methods can be found in the literature, see \cite{key-1-1-1-1-8,key-1-1-1-1-9,key-1-1-1-1-10,key-1-1-1-1-10-1}
for an overview.

\noindent Regression methods have also been applied to adjust for
publication bias. Moreno et al. (2011) regresses the treatment effect
on its variance, using the variance as a weight, and uses the predicted
value for an infinitely large study as an adjusted estimate for the
treatment effect \cite{key-1-3}. Moreno's method is related to the
PET-PEESE approach in which Stanley and Doucoullagos (2007) suggest
using the slope estimate from the Egger model as bias-adjusted overall
treatment effect estimate, if the null hypothesis of zero treatment
effect is not rejected \cite{key-1-1-5,key-1-1-6}. If this null hypothesis
is rejected, the method uses the intercept estimate from Moreno's
regression equation as a bias-adjusted overall estimate. Although
the PET-PEESE method was shown to perform well with continuous outcomes,
its estimate was found to be less efficient when there is treatment
effects heterogeneity \cite{key-1-6}. Rucker et al. (2011) developed
the limit meta-analysis method to adjust for publication bias. The
authors adjust the random-effect meta-analysis model by introducing
a publication bias parameter, and suggest the Maximum Likelihood method
to estimate the model parameters. Noting that the estimates can be
obtained using a linear regression model on the so-called generalized
radial plot, with the standardized effect size dependent on the inverse
of the standard error \cite{key-1-1-1-9}. 

\noindent Comparison studies of methods adjusting for publication
bias are found in the literature. Terrin et al. (2003) compared the
performance of the Trim and Fill method with that of Hedges selection
model in the case of odds ratio \cite{key-1-1-1-1-1-4}. The authors
concluded that Hedges method performed better than the Trim \& Fill
method, but noted the convergence problems of the iterative Hedges
method. Schwarzer et al. (2010) compared the performance of the Trim
and Fill method with that of the Copas method in case of binary outcomes
using the arcsine difference as an effect size. The authors concluded
that the Trim and Fill method produced bigger standard error, wider
confidence intervals, i.e. it is more conservative and less efficient
than the Copas method \cite{key-1-1-1-1-12}. Reed (2015) compared
six meta-analysis estimators in case of publication bias using a simulation
study \cite{key-1-1-20}. Publication bias was created based on two
scenarios: bias towards statistically significant estimates and bias
against studies with wrong-signed estimated effects. The simulation
set-up used is suitable for both binary and continuous outcomes. The
six estimators considered were: the fixed-effects estimate, the weighted
least squares estimator, the random-effects estimator, the PET estimator
\cite{key-1-1-5,key-1-1-6}, the PEESE estimator \cite{key-1-1-5,key-1-1-6},
while using the arithmetic mean of estimated effects as the benchmark
estimator. Reed concluded that the fixed-effects and the weighted
least squares estimators are as efficient and sometimes more efficient
than the PET and the PEESE estimators, while the random-effects estimator
is often more biased than the other estimators \cite{key-1-1-20}.

\noindent Mcshane et al. (2016) compared the performance of a number
of publication bias adjustment methods using the mean difference as
an effect size and applying one-sided selection \cite{key-1-1-21}.
The authors compared the p-curve and the p-uniform approaches to the
maximum likelihood estimation approaches of Hedges (1984) and Iyengar
and Greenhouse (1988) \cite{key-1-1-1-1-1}. The p-curve and p-uniform
approaches were found to perform well under the setting they were
designed for: namely when only studies with results that are statistically
significant and directionally consistent are published and effect
sizes are homogeneous across studies. However, when one or both of
these assumptions is violated, the p-curve and the p-uniform methods
become less accurate than the Iyengar and Greenhouse method. Although
McShane et al. (2016) compares the performance of a few methods in
case of continuous outcomes, the authors did not consider other methods
not based on selection models and studied only one type of effect
size. 

\noindent This brief review exposes the lack of a comprehensive comparative
study for methods adjusting for publication bias in case of continuous
outcomes. Investigating all adjustment methods in the literature is
obviously beyond the scope of this article. We selected five publication
bias adjustment methods trying to represent different techniques applied
by the methods, and applied these methods in the case of continuous
outcomes using Cohen\textquoteright s d and Hedges\textquoteright{}
g as treatment effect estimates. One popular method based on selection
models, the Copas method, is considered. Another method based on the
p-values, the p uniform method, is considered. The performance of
the PET-PEESE method, a regression analysis based method, is investigated.
Finally, two Likelihood-based methods the Trim \& Fill method and
the limit meta-analysis method are considered. In addition, the traditional
random effect meta-analysis method based on the DerSimonian-Laird
estimate is also included. \foreignlanguage{british}{The layout of
this article is as follows. In section 2.2 we describe the selected
methods used to adjust publication bias}. Section 3 describes the
simulation model and the selection model used to simulate publication
bias. The results and the discussion are relegated to sections 4 and
5, respectively.

\section{Adjustment methods}

A meta-analysis of continuous outcomes usually consists of\foreignlanguage{british}{
$m$ studies, $i$ ($i=1,2,...,m$). Each study typically has $n_{i}$
participants, with $n_{ij}$ subjects assigned to treatment arm $j,$
$j\epsilon\left\{ 0,1\right\} $. Here $j=1$ and $j=0$ indicate
the treatment and the control arms, respectively. }Define $\hat{y}_{i1}$
and $S_{i1}$ as the average and the standard deviation of the treatment
arm, respectively, and $\hat{y}_{i0}$ and $S_{i0}$ the average and
the standard deviation of the control arm, respectively. Traditionally
the mean difference \foreignlanguage{british}{$\hat{y}_{i1}-\hat{y}_{i0}$
}is used as a treatment effect for study $i$. Another treatment effect
measure is Cohen's d which can be given by $\left(\hat{y}_{i1}-\hat{y}_{i0}\right)/se_{i}$
where $se_{i}=\sqrt{\left(\left(n_{i1}-1\right)*S_{i1}^{2}+\left(n_{i0}-1\right)*S_{i0}^{2}\right)/\left(n_{i1}+n_{i0}-2\right)}.$\foreignlanguage{british}{
However, Cohen's $d$ has been found to be biased \cite{key-1-1-1-1-2-1}
and a corrected unbiased estimate is referred to as Hedges' $g$.
It multiplies Cohen's $d$ with the correction factor $J=1-3/\left(4\left(n_{1}+n_{0}-2\right)-1\right)$.
From here on we refer to the treatment effect for study $i$ by }$\hat{y}_{i}$
which can refer to both Cohen's d and Hedges g.

\subsection{Copas selection model}

Copas and Shi \cite{key-1-1-1-1-4,key-1-1-1-1-5} assume that all
studies follow the traditional random-effects meta-analysis model
given by

\begin{equation}
y_{i}=y+\theta_{i}+\epsilon_{i}\label{randomeffectsmodel}
\end{equation}

\noindent with $y$ general treatment effect, $\theta_{i}\sim N\left(0,\tau^{2}\right)$
and $\epsilon_{i}\sim N\left(0,V_{i}^{2}\right)$. However, the authors
assume that only a subset of the studies has been published. The authors
then introduce $z_{i}=A_{1}+A_{2}/\hat{se_{i}}+\delta_{i}$, with
$\delta_{i}\sim N(0,1)$ correlated with $\epsilon_{i}$ through $\rho=corr(\epsilon_{i},\delta_{i})$,
and $\hat{y}_{i}$ only reported if $z_{i}>0$. The conditional probability
density function will be given by $P\left(\hat{y}_{i}|z_{i}>0,\hat{se_{i}}\right)$
and the log Likelihood function is constructed as

\noindent
\[
L\left(y,\tau^{2}\right)=\sum_{i=1}^{m}\left[\log P\left(\hat{y}_{i}|z_{i}>0,\hat{se_{i}}\right)\right].
\]

\noindent For fixed values of $A_{1}$ and $A_{2}$ , $L\left(y,\tau^{2}\right)$
is maximized and $y$ and $\tau^{2}$ along with their confidence
interval are estimated. We used the R-package ``copas'' which is
part of the R-package meta to carry out the Copas method \cite{key-1-1-1-1-4,key-1-1-1-1-5}. 

\subsection{The p-uniform method}

\noindent The p-uniform is a selection model approach, with the selection
model assuming that the probability of publishing a statistically
significant treatment effect as well as a non significant treatment
effect are constant but may differ from each other \cite{key-1-1-8}.
Defining the $F_{X}\left(x_{i}\right)$ the distribution functions
of the normal distributions $N\left(y^{f},\hat{se_{i}}\right)$, with
$y^{f}$ the fixed-effects meta analysis effect size estimate and
$\dot{y_{i}}$ the critical value for study $i$, the authors define
the conditional p value distribution

\[
p_{i}^{\dot{y}}=\frac{\intop_{\hat{y}_{i}}^{\infty}F_{X}\left(x_{i}\right)}{\intop_{\dot{y_{i}}}^{\infty}F_{X}\left(x_{i}\right)}
\]

\noindent where $\dot{y}$ is the critical value for rejecting the
null hypothesis of no treatment effect at significance level $\alpha$.
Using the gamma distributed test statistic $L^{y}=-\sum_{i=1}^{m}\log\left(p_{i}^{y}\right)$,
the authors introduce the bias-adjusted estimate $\hat{y}$ for $y$,
with $\hat{y}$ the solution to $L^{\hat{y}}=m$. The p-uniform lower
and upper limits of the point estimate $\hat{y}_{L}$ and $\hat{y}_{U}$
are then given by the solutions of $L^{\hat{y}_{L}}=\Gamma_{1-\alpha/2}\left(m,1\right)$
and $L^{\hat{y}_{U}}=\Gamma_{\alpha/2}\left(m,1\right)$, respectively,
with $\Gamma$ the gamma function. We used the package \textquoteleft puniform\textquoteright{}
to perform the p-uniform method \cite{key-1-1-9}.

\subsection{The PET-PEESE method}

Stanley and Doucouliagos (2013) start with the linear regression model
used in the Egger test \cite{key-1-1-5,key-1-1-6} given by

\begin{equation}
\frac{\hat{y}_{i}}{\hat{se_{i}}}=\alpha_{0}+\frac{\alpha_{1}}{\hat{se_{i}}}+\varepsilon_{i},\label{eq:EGGER=000020TEST}
\end{equation}

\noindent with $\alpha_{0}$ and $\alpha_{1}$ an intercept and a
regression coefficient, respectively, and $\varepsilon_{i}\sim N\left(0,\sigma^{2}\right)$.
The authors suggest using $\hat{\alpha}_{1}$ in model (\ref{eq:EGGER=000020TEST})
as a bias-adjusted estimate for the overall treatment effect, if the
null hypothesis of $\alpha_{1}=0$ is not rejected by the Egger equation
using a t-test (PET). If this null hypothesis is rejected, the model
$y_{i}=\gamma_{0}+\gamma_{1}se_{i}^{2}+\epsilon_{i}$, $\epsilon_{i}\sim N\left(0,\sigma^{2}\right)$
is fitted using $1/se_{i}^{2}$ as weight, and $\hat{\gamma}_{0}$
is used as a bias-adjusted estimate of the overall treatment effect
(PEESE). Note that the last regression model was the one suggested
by Moreno et al. (2011). We used Procedure GLM in SAS to carry out
the PET-PEESE method.

\subsection{Trim \& Fill Method}

The Trim \& fill method can be briefly described as follows \cite{key-1-1-1-1-1-2,key-1-1-1-1-1-3}.
The studies are first ranked based on their distance from the pooled
treatment effect estimated by the random effects model $\hat{y}_{RM}$
(\ref{randomeffectsmodel}), i.e., ranking distances $|\hat{y}_{i}-\hat{y}_{RM}|$.
Next, the number of unobserved studies is estimated using the estimator
$L_{0}=[4T_{m}-m(m+1)]/[2m-1]$, where $T_{m}$ is the Wilcoxon rank
sum test statistic estimated from the ranks of studies with $\hat{y}_{i}>\hat{y}_{RM}$
(here it is assumed that $\hat{y}_{RM}$ is positive and it is more
likely that studies with effect sizes below $\hat{y}_{RM}$ are potentially
missing). Then the $L_{0}$ most extreme studies (i.e., studies with
positive effect sizes furthest away from zero) are trimmed off, and
the pooled treatment effect $\hat{y}_{RM}$ without these studies
is re-estimated. Then all studies are ranked again, based on their
distance to the new pooled estimate, and $L_{0}$ is re-computed.
This procedure is repeated until it stabilizes ($L_{0}$ does not
change anymore) and we obtain a final estimate $\hat{y}_{RM}$ and
a final estimate $L_{0}$ of the number of studies missing. Then $L_{0}$
studies are imputed by mirroring the $L_{0}$ studies with the highest
effect sizes around the final estimate for $\hat{y}_{RM}$ and the
standard error $se_{i}$ from the mirrored study is provided. After
imputation, a final pooled estimate $\hat{y}_{TF}$ with standard
error is provided using the random effects model on all $m+L_{0}$
studies. We used the function ``trimfill'' in the R package ``metafor'',
with the average treatment effect estimated using the random effects
model \cite{key-1-1-7}.

\subsection{The limit meta-analysis method}

The limit meta-analysis can be described as follows \cite{key-1-1-1-9}.
The authors apply a publication bias adjusted version of the random-effects
meta-analysis model as follows

\[
\hat{y}_{i}=y+\left(se_{i}+\tau\right)\left(\epsilon_{i}+\alpha\right),
\]

\noindent with $\alpha$ a parameter representing publication bias.
The authors proposed using $\hat{y}+\hat{\tau}\hat{\alpha}$ as a
publication bias adjusted estimator for the effect size $y$. The
authors proposed using the Maximum Likelihood method to estimate the
parameters $y$ and $\tau$. However, the authors noted that Maximum
Likelihood estimates for $y$ and $\tau$ correspond with the intercept
and slope in linear regression on a generalized radial plot. In such
model the response variable and the independent variable are given
by $\hat{y}_{i}/\left(\hat{se_{i}}+\hat{\tau}\right)$ and $1/\left(\hat{se_{i}}+\hat{\tau}\right)$,
respectively \cite{key-1-1-1-9}. We used the R package 'metasens'
to carry out the limit meta-analysis method \cite{key-1-1-1-9-1}. 

\section{Case study, Simulation and selection models}

\subsection{Case study}

As a case study we used the results of 25 studies investigating the
effect of the experience of weight on the concept of importance. Experience
of weight, exemplified by heaviness and lightness, is metaphorically
associated with concepts of seriousness and importance \cite{key-1-2-1-1}.
Robelo et al. (2015) reported 25 studies that investigated the effect
of weight on the concept of importance. All published studies had
continuous outcomes for the expression of importance and all applied
Cohen's d as an effect size \cite{key-1-2-1-2}. There was evidence
that these published studies were too good to be true \cite{key-1-2-1-3}.
In addition, Rabelo et al. (2015) conducted their own research on
this topic after applying p-uniform's test for publication bias on
these published studies \cite{key-1-1-1-1-7}, and finding evidence
of significant publication bias \cite{key-1-2-1-2}. The dataset containing
the results of the 25 studies can found in Rabelo et al. (2015) \cite{key-1-2-1-2}.

\subsection{Simulation model}

The simulation model used is described in subsection 3.2.1. The simulation
model explores different scenario's including different study sizes,
number of units per treatment arm per study, treatment effect homogeneity
and different levels of treatment effect heterogeneity. In addition,
two scenario's of equal and unequal study variance scenario's for
the treatment and control groups, respectively. A separte model for
selecting a number of studies based on specific criteria to simulate
publication bias is described in subsection 3.2.2. 

\subsubsection{Aggregate simulation model}

The simulation model applied can be described as follows \cite{key-1-1-1-1-14,key-1-1-1-1-15}.
In total $m$ studies are generated, and for the $i^{th}$ study $n_{ij}$,
$j=0,1$, are independently drawn from a Poisson distribution with
parameter $\delta$. The estimated treatment effect for the $j^{th}$
treatment group in study $i$ is given by $\overline{y}_{ij}=\left(\eta+\theta_{i}\right)\cdotp t_{ij}+\varepsilon_{ij}$,
with $\eta$ a fixed treatment effect, $\theta_{i}$ a random effect
for study $i$, $\theta_{i}\sim N\left(0,\tau^{2}\right)$, $t_{ij}$
a treatment indicator variable for study $i$ with value 1 when $j=1$
and $0$ otherwise, and $\varepsilon_{ij}$ an error term with $\varepsilon_{ij}\sim N\left(0,S_{ij}^{2}/n_{ij}\right)$.
For the equal study variance scenario $S_{i0}^{2}=S_{i1}^{2}=S_{i}^{2}$,
while for the unequal study variance scenario the variance for the
control and the treatment groups are created by $S_{i0}^{2}=0.8S_{i}^{2}$
and $S_{i1}^{2}=S_{i}^{2}$, respectively. The parameter values used
to generate the data are $m\in\{10,30\}$, $\delta=\left\{ 15,30\right\} $,
$\eta=5$, $\tau^{2}=\left\{ 0,1,5\right\} $, and $S_{i}^{2}$ is
generated using $S_{i}^{2}\sim N\left(100,100\right)$. Cohen's d
and Hedges g are then calculated as described in section 2 earlier. 

\subsubsection{Selection model based on significant effect size}

Selection models based on the $p$-value of the study effect have
been proposed in literature \cite{key-1-1-5,key-1-1-6,key-1-1-21,key-1-1-9}.
Let $\alpha$ the significance level and $t_{q,\nu_{i},}$ the $q^{th}$\textsuperscript{}
quantile of a student t distribution with $\nu_{i}$ degrees of freedom
for study $i$. When the effect size is significant (assuming more
positive effect sizes), i.e., $d_{i}>t_{\left(1-\alpha/2\right),\nu_{i}}$,
the study is included. To add randomness to the selected studies,
a uniform distributed random variable $U(0,1)$ and a parameter $\pi_{\mathrm{pub}}$
are used. If the uniform random variable is smaller than or equal
to $1-\pi_{\mathrm{pub}}$, the non-significant study is included
too; otherwise it is excluded. Here $\alpha=0.05$ and $\pi_{\mathrm{pub}}$
is chosen such that desired publishing rate of around 80\% is obtained.
For the equal study variance scenario we applied $\nu=n_{i0}+n_{i1}-2$.
For the unequal variance scenario we applied the Satterthwaite approximation
for the degrees of freedom $\nu=\left(S_{ic}^{2}/n_{ic}+S_{iT}^{2}/n_{iT}\right)^{2}/\left(\left(S_{ic}^{2}/n_{ic}\right)^{2}/\left(n_{ic}-1\right)+\left(S_{iT}^{2}/n_{iT}\right)^{2}/\left(n_{iT}-1\right)\right)$
\cite{key-1-1-1-1-17}.

\section{Results}

\subsection{Results for the case study}

Table 1 shows the treatment effect estimates for the case study described
in section 3.1 using the five adjustment methods explained in section
2. Also presented is the treatment effect estimate using the traditional
random effects meta analysis method based on the DerSimonian-Laird
estimator (DL). All methods, with the exception of the PET-PEESE and
the limit meta methods, produce a treatment effect estimate of between
0.5 and 0.6. The limit meta method produces a lower treatment effect
estimate, while the PET-PEESE is the only method which indicates that
there is no treatment effect.

\medskip{}

\begin{table}[H]

\caption{Results of treatment effect estimates for the case study}

\begin{tabular}{|c|c|c|}
\hline 
Method & Cohen's d & Hedges g\tabularnewline
\hline 
\hline 
Copas & 0.54665 (0.44416; 0.64914) & 0.57052 (0.46803 ; 0.67301)\tabularnewline
\hline 
P-uniform & 0.4786477 (-2.560318 ; 0.7836985) & 0.4597283 (-3.244159 ; 0.7352272 )\tabularnewline
\hline 
PET-PEESE & -0.001796206 (-0.034486 ; 0.030893) & -0.001712641 (-0.033384 ; 0.029959)\tabularnewline
\hline 
Trim \& Fill & 0.54305 (0.36237 ; 0.72373) & 0.54064 (0.35672 ; 0.72455)\tabularnewline
\hline 
Limit meta & 0.3135616 (-0.5641693 ; 0.6649695) & 0.3135616 (-0.548716 ; 0.6804229)\tabularnewline
\hline 
DL & 0.57825 (0.47577 ; 0.68074) & 0.57052 (0.46803 ; 0.67301)\tabularnewline
\hline 
\end{tabular}

\end{table}

\subsection{Results for the simulation study}

\subsubsection{Average number of studies after selection}

Table 2 shows the average number of remaining studies per simulated
meta-analysis after applying the selection model explained in subsection
3.2.2 for all simulation settings. We tried to keep the average number
of remaining studies around 80\% of original studies for comparison
purposes. This percentage is based on earlier research concluding
that the reported proportion of published studies with positive outcomes
was 82\% in emergency or general medicine \cite{key-1-1-1-1-16}.

\medskip{}

\begin{table}[H]
\caption{Average number of remaining studies after applying selection model
for the different simulation settings}

\medskip{}

\begin{tabular*}{1cm}{@{\extracolsep{\fill}}|>{\raggedright}p{1cm}|>{\raggedright}p{1cm}|>{\raggedright}p{1cm}|>{\raggedright}p{1cm}|>{\raggedright}p{1cm}|>{\raggedright}p{1cm}|>{\raggedright}p{1cm}|}
\hline 
\multirow{2}{1cm}{$m$} & \multirow{2}{1cm}{$n_{i}$} & \multirow{2}{1cm}{$\tau^{2}$} & \multicolumn{2}{c|}{Cohen's d} & \multicolumn{2}{c|}{Hedges g}\tabularnewline
\cline{4-7}
 &  &  & Equal variance & Unequal variance & Equal variance & Unequal variance\tabularnewline
\hline 
\multirow{6}{1cm}{10} & \multirow{3}{1cm}{15} & \multirow{1}{1cm}{0} & 8.23  & 8.09 & 8.04 & 8.08\tabularnewline
\cline{3-7}
 &  & \multirow{1}{1cm}{1} & 8.22 & 8.08 & 8.02 & 8.07\tabularnewline
\cline{3-7}
 &  & \multirow{1}{1cm}{5} & 8.21 & 8.01 & 8.01 & 7.98\tabularnewline
\cline{2-7}
 & \multirow{3}{1cm}{30} & \multirow{1}{1cm}{0} & 8.11 & 8.09 & 8.20 & 8.00\tabularnewline
\cline{3-7}
 &  & \multirow{1}{1cm}{1} & 8.13 & 8.10 & 7.99 & 7.98\tabularnewline
\cline{3-7}
 &  & \multirow{1}{1cm}{5} & 7.87 & 8.14 & 8.12 & 8.12\tabularnewline
\hline 
\multirow{6}{1cm}{30} & \multirow{3}{1cm}{15} & \multirow{1}{1cm}{0} & 23.0 & 24.1 & 24.1 & 24.0\tabularnewline
\cline{3-7}
 &  & \multirow{1}{1cm}{1} & 23.0 & 24.1 & 24.1 & 24.0\tabularnewline
\cline{3-7}
 &  & \multirow{1}{1cm}{5} & 24.5 & 24.0 & 24.1 & 23.9\tabularnewline
\cline{2-7}
 & \multirow{3}{1cm}{30} & \multirow{1}{1cm}{0} & 24.2 & 23.9 & 24.2 & 23.9\tabularnewline
\cline{3-7}
 &  & \multirow{1}{1cm}{1} & 24.2 & 24.1 & 23.7 & 23.8\tabularnewline
\cline{3-7}
 &  & \multirow{1}{1cm}{5} & 23.4 & 24.1 & 23.9 & 24.1\tabularnewline
\hline 
\end{tabular*}
\end{table}

\subsubsection{Copas selection model}

For Cohen's d in case of $m=10$ and $m=30$, the AMSE and bias of
the Copas method are higher for the unequal study variance scenario
when $n_{i}=15$ but not always when $n_{i}=30$. The coverage probability
for Cohen's d is comparable for the equal and unequal study variance
scenario's. However, the coverage probability decreases when $m=30$
for both homogeneity and heterogeneity of the treatment effect. Hedges
g shows comparable values of AMSE, bias and coverage probability when
$m=1$0 and $m=30$ when $n_{i}=10$ for different values of $\tau^{2}$
and for both the equal and unequal study variance scenario's. However,
the AMSE, bias and coverage probability clearly decrease when $n_{i}=30$.
Overall, Hedges g has a smaller AMSE and bias and higher coverage
probabilities than Cohen's d. The AMSE and bias of the method under
the equal variance scenario are lower than under the unequal variance
scenario. The statistical power of the Copas method is lower under
the equal variance scenario.

\medskip{}

\begin{table}[H]
\caption{An overview of the AMSE, bias and coverage probabilities for the Copas
method under equal variance assumption for the different simulation
settings.}

\medskip{}
\begin{tabular*}{1cm}{@{\extracolsep{\fill}}|>{\raggedright}p{1cm}|>{\raggedright}p{1cm}|>{\raggedright}p{1cm}|>{\raggedright}p{1cm}|>{\raggedright}p{1cm}|>{\raggedright}p{1cm}|>{\raggedright}p{1cm}|>{\raggedright}p{1cm}|>{\raggedright}p{1cm}|}
\hline 
\multirow{2}{1cm}{$m$} & \multirow{2}{1cm}{$n_{i}$} & \multirow{2}{1cm}{$\tau^{2}$} & \multicolumn{3}{c|}{Cohen D} & \multicolumn{3}{c|}{Hedges G}\tabularnewline
\cline{4-9}
 &  &  & \multicolumn{1}{l|}{Amse} & \multicolumn{1}{l|}{Bias} & \multicolumn{1}{l|}{COV} & \multicolumn{1}{l|}{Amse} & \multicolumn{1}{l|}{Bias} & \multicolumn{1}{l|}{COV}\tabularnewline
\hline 
\multirow{6}{1cm}{10} & \multirow{3}{1cm}{15} & \multirow{1}{1cm}{0} & 1.54 & 0.344 & 0.767 & 1.55 & 0.376 & 0.755\tabularnewline
\cline{3-9}
 &  & \multirow{1}{1cm}{1} & 1.60 & 0.351 & 0.784 & 1.65 & 0.371 & 0.779\tabularnewline
\cline{3-9}
 &  & \multirow{1}{1cm}{5} & 1.13 & 0.424 & 0.754 & 1.66 & 0.346 & 0.764\tabularnewline
\cline{2-9}
 & \multirow{3}{1cm}{30} & \multirow{1}{1cm}{0} & 0.927 & 0.232 & 0.587 & 0.944 & 0.227 & 0.597\tabularnewline
\cline{3-9}
 &  & \multirow{1}{1cm}{1} & 0.928 & 0.226 & 0.586 & 0.974 & 0.285 & 0.584\tabularnewline
\cline{3-9}
 &  & \multirow{1}{1cm}{5} & 1.37 & 0.390 & 0.666 & 1.37 & 0.325 & 0.703\tabularnewline
\hline 
\multirow{6}{1cm}{30} & \multirow{3}{1cm}{15} & \multirow{1}{1cm}{0} & 1.11 & 0.420 & 0.763 & 0.895 & 0.321 & 0.801\tabularnewline
\cline{3-9}
 &  & \multirow{1}{1cm}{1} & 1.13 & 0.424 & 0.754 & 0.906 & 0.324 & 0.797\tabularnewline
\cline{3-9}
 &  & \multirow{1}{1cm}{5} & 1.04 & 0.372 & 0.842 & 1.021 & 0.384 & 0.833\tabularnewline
\cline{2-9}
 & \multirow{3}{1cm}{30} & \multirow{1}{1cm}{0} & 0.520 & -0.146 & 0.651 & 0.503 & -0.138 & 0.657\tabularnewline
\cline{3-9}
 &  & \multirow{1}{1cm}{1} & 0.540 & -0.137 & 0.675 & 0.505 & -0.076 & 0.677\tabularnewline
\cline{3-9}
 &  & \multirow{1}{1cm}{5} & 0.877 & 0.185 & 0.761 & 0.807 & 0.157 & 0.766\tabularnewline
\hline 
\end{tabular*}
\end{table}

\medskip{}
\begin{table}[H]
\caption{An overview of the AMSE, bias and coverage probabilities for the Copas
method under unequal variance assumption for the different simulation
settings.}

\medskip{}
\begin{tabular*}{1cm}{@{\extracolsep{\fill}}|>{\raggedright}p{1cm}|>{\raggedright}p{1cm}|>{\raggedright}p{1cm}|>{\raggedright}p{1cm}|>{\raggedright}p{1cm}|>{\raggedright}p{1cm}|>{\raggedright}p{1cm}|>{\raggedright}p{1cm}|>{\raggedright}p{1cm}|}
\hline 
\multirow{2}{1cm}{$m$} & \multirow{2}{1cm}{$n_{i}$} & \multirow{2}{1cm}{$\tau^{2}$} & \multicolumn{3}{c|}{Cohen D} & \multicolumn{3}{c|}{Hedges G}\tabularnewline
\cline{4-9}
 &  &  & \multicolumn{1}{l|}{Amse} & \multicolumn{1}{l|}{Bias} & \multicolumn{1}{l|}{COV} & \multicolumn{1}{l|}{Amse} & \multicolumn{1}{l|}{Bias} & \multicolumn{1}{l|}{COV}\tabularnewline
\hline 
\multirow{6}{1cm}{10} & \multirow{3}{1cm}{15} & \multirow{1}{1cm}{0} & 2.18 & 0.846 & \multirow{1}{1cm}{0.656} & 2.10 & 0.843 & 0.653\tabularnewline
\cline{3-9}
 &  & \multirow{1}{1cm}{1} & 2.15 & 0.842 & 0.661 & 2.07 & 0.837 & 0.647\tabularnewline
\cline{3-9}
 &  & \multirow{1}{1cm}{5} & 2.44 & 0.830 & 0.692 & 2.32 & 0.843 & 0.673\tabularnewline
\cline{2-9}
 & \multirow{3}{1cm}{30} & \multirow{1}{1cm}{0} & 1.53 & 0.831 & 0.390 & 1.62 & 0.866 & 0.364\tabularnewline
\cline{3-9}
 &  & \multirow{1}{1cm}{1} & 1.52 & 0.840 & 0.378 & 1.59 & 0.886 & 0.363\tabularnewline
\cline{3-9}
 &  & \multirow{1}{1cm}{5} & 1.85 & 0.836 & 0.587 & 1.84 & 0.840 & 0.590\tabularnewline
\hline 
\multirow{6}{1cm}{30} & \multirow{3}{1cm}{15} & \multirow{1}{1cm}{0} & 1.47 & 0.795 & 0.703 & 1.39 & 0.781 & 0.715\tabularnewline
\cline{3-9}
 &  & \multirow{1}{1cm}{1} & 1.49 & 0.795 & 0.688 & 1.41 & 0.781 & 0.711\tabularnewline
\cline{3-9}
 &  & \multirow{1}{1cm}{5} & 1.74 & 0.846 & 0.689 & 1.64 & 0.834 & 0.695\tabularnewline
\cline{2-9}
 & \multirow{3}{1cm}{30} & \multirow{1}{1cm}{0} & 0.640 & 0.425 & 0.566 & 0.622 & 0.424 & 0.560\tabularnewline
\cline{3-9}
 &  & \multirow{1}{1cm}{1} & 0.676 & 0.431 & 0.564 & 0.645 & 0.432 & 0.565\tabularnewline
\cline{3-9}
 &  & \multirow{1}{1cm}{5} & 1.17 & 0.597 & 0.697 & 1.15 & 0.604 & 0.702\tabularnewline
\hline 
\end{tabular*}
\end{table}

\subsubsection{The p-uniform method}

For Cohen's d the AMSE and bias of the p-uniform method are comparable
for the unequal study variance scenario for all simulation settings.
The coverage probability mostly increases for Cohen's d as the treatment
effect heterogeneity increases. The AMSE and bias of the Cohen's D
decrease while the coverage probability increases as the treatment
effect heterogeneity increases. Hedges g shows comparable values of
AMSE, bias and coverage probability when $n_{i}=15$ for different
values of $\tau^{2}$ and for both the equal and unequal study variance
scenario's. However, the AMSE, bias and coverage probability mostly
increase when $n_{i}=30$, except when $\tau^{2}$=5. The coverage
probability mostly increases for Hedges G as the treatment effect
heterogeneity increases. Cohen's d and Hedges g have comparable bias,
AMSE and coverage probabilities. The AMSE and bias of the p-uniform
method under the equal variance scenario are lower than under the
unequal variance scenario. The statistical power of the method under
the equal and the unequal variance scenario are comparable.

\medskip{}
\begin{table}[H]
\caption{An overview of the AMSE, bias and coverage probabilities for the p-uniform
method under equal variance assumption for the different simulation
settings.}

\medskip{}
\begin{tabular*}{1cm}{@{\extracolsep{\fill}}|>{\raggedright}m{1cm}|>{\raggedright}m{1cm}|>{\raggedright}m{1cm}|>{\raggedright}m{1cm}|>{\raggedright}m{1cm}|>{\raggedright}m{1cm}|>{\raggedright}m{1cm}|>{\raggedright}m{1cm}|>{\raggedright}m{1cm}|}
\hline 
\multirow{2}{1cm}{$m$} & \multirow{2}{1cm}{$n_{i}$} & \multirow{2}{1cm}{$\tau^{2}$} & \multicolumn{3}{c|}{Cohen D} & \multicolumn{3}{c|}{Hedges G}\tabularnewline
\cline{4-9}
 &  &  & \multicolumn{1}{l|}{Amse} & \multicolumn{1}{l|}{Bias} & \multicolumn{1}{l|}{COV} & \multicolumn{1}{l|}{Amse} & \multicolumn{1}{l|}{Bias} & \multicolumn{1}{l|}{COV}\tabularnewline
\hline 
\multirow{7}{1cm}{10} & \multirow{4}{1cm}{15} & \multirow{2}{1cm}{0} & \multirow{2}{1cm}{14.2} & \multirow{2}{1cm}{3.58} & \multirow{2}{1cm}{0.007} & \multirow{2}{1cm}{13.6} & \multirow{2}{1cm}{3.51} & \multirow{2}{1cm}{0.009}\tabularnewline
 &  &  &  &  &  &  &  & \tabularnewline
\cline{3-9}
 &  & \multirow{1}{1cm}{1} & 14.3 & 3.60 & 0.007 & 13.8 & 3.53 & 0.006\tabularnewline
\cline{3-9}
 &  & \multirow{1}{1cm}{5} & 13.2 & 3.57 & <0.001  & 16.2 & 3.84 & 0.003\tabularnewline
\cline{2-9}
 & \multirow{3}{1cm}{30} & \multirow{1}{1cm}{0} & 11.2 & 3.24 & 0.002 & 10.8 & 3.18 & 0.001\tabularnewline
\cline{3-9}
 &  & \multirow{1}{1cm}{1} & 11.2 & 3.24 & 0.001 & 11.0 & 3.22 & 0.001\tabularnewline
\cline{3-9}
 &  & \multirow{1}{1cm}{5} & 16.9 & 3.93 & 0.003 & 16.0 & 3.78 & 0.009\tabularnewline
\hline 
\multirow{6}{1cm}{30} & \multirow{3}{1cm}{15} & \multirow{1}{1cm}{0} & 13.1 & 3.56 & <0.001  & 12.2 & 3.42 & <0.001 \tabularnewline
\cline{3-9}
 &  & \multirow{1}{1cm}{1} & 13.2 & 3.57 & <0.001  & 12.2 & 3.43 & <0.001 \tabularnewline
\cline{3-9}
 &  & \multirow{1}{1cm}{5} & 15.3 & 3.84 & <0.001  & 14.7 & 3.76 & <0.001 \tabularnewline
\cline{2-9}
 & \multirow{3}{1cm}{30} & \multirow{1}{1cm}{0} & 10.4 & 3.19 & <0.001  & 10.2 & 3.15 & <0.001 \tabularnewline
\cline{3-9}
 &  & \multirow{1}{1cm}{1} & 10.6 & 3.21 & <0.001  & 10.5 & 3.19 & <0.001 \tabularnewline
\cline{3-9}
 &  & \multirow{1}{1cm}{5} & 15.9 & 3.92 & <0.001  & 15.3 & 3.85 & <0.001 \tabularnewline
\hline 
\end{tabular*}
\end{table}

\medskip{}

\begin{table}[H]
\caption{An overview of the AMSE, bias and coverage probabilities for the p-uniform
method under unequal variance assumption for the different simulation
settings.}

\medskip{}
\begin{tabular*}{1cm}{@{\extracolsep{\fill}}|>{\raggedright}p{1cm}|>{\raggedright}p{1cm}|>{\raggedright}p{1cm}|>{\raggedright}p{1cm}|>{\raggedright}p{1cm}|>{\raggedright}p{1cm}|>{\raggedright}p{1cm}|>{\raggedright}p{1cm}|>{\raggedright}p{1cm}|}
\hline 
\multirow{2}{1cm}{$m$} & \multirow{2}{1cm}{$n_{i}$} & \multirow{2}{1cm}{$\tau^{2}$} & \multicolumn{3}{c|}{Cohen D} & \multicolumn{3}{c|}{Hedges G}\tabularnewline
\cline{4-9}
 &  &  & \multicolumn{1}{l|}{Amse} & \multicolumn{1}{l|}{Bias} & \multicolumn{1}{l|}{COV} & \multicolumn{1}{l|}{Amse} & \multicolumn{1}{l|}{Bias} & \multicolumn{1}{l|}{COV}\tabularnewline
\hline 
\multirow{7}{1cm}{10} & \multirow{4}{1cm}{15} & \multirow{2}{1cm}{0} & \multirow{2}{1cm}{17.7} & \multirow{2}{1cm}{4.05} & \multirow{2}{1cm}{0.003} & \multirow{2}{1cm}{16.9} & \multirow{2}{1cm}{3.96} & \multirow{2}{1cm}{0.003}\tabularnewline
 &  &  &  &  &  &  &  & \tabularnewline
\cline{3-9}
 &  & \multirow{1}{1cm}{1} & 17.7 & 4.05 & 0.002 & 16.9 & 3.96 & 0.004\tabularnewline
\cline{3-9}
 &  & \multirow{1}{1cm}{5} & 21.3 & 4.41 & 0.00400 & 20.4 & 4.31 & 0.002\tabularnewline
\cline{2-9}
 & \multirow{3}{1cm}{30} & \multirow{1}{1cm}{0} & 14.9 & 3.78 & 0.00100 & 14.7 & 3.75 & 0.002\tabularnewline
\cline{3-9}
 &  & \multirow{1}{1cm}{1} & 15.1 & 3.80 & <0.001  & 14.9 & 3.77 & <0.001 \tabularnewline
\cline{3-9}
 &  & \multirow{1}{1cm}{5} & 21.6 & 4.48 & 0.001 & 21.2 & 4.44 & <0.001 \tabularnewline
\hline 
\multirow{6}{1cm}{30} & \multirow{3}{1cm}{15} & \multirow{1}{1cm}{0} & 16.0 & 3.94 & <0.001  & 15.3 & 3.86 & <0.001 \tabularnewline
\cline{3-9}
 &  & \multirow{1}{1cm}{1} & 16.0 & 3.95 & <0.001  & 15.4 & 3.86 & <0.001 \tabularnewline
\cline{3-9}
 &  & \multirow{1}{1cm}{5} & 19.0 & 4.29 & <0.001  & 18.2 & 4.21 & <0.001 \tabularnewline
\cline{2-9}
 & \multirow{3}{1cm}{30} & \multirow{1}{1cm}{0} & 14.2 & 3.73 & <0.001  & 13.9 & 3.69 & <0.001 \tabularnewline
\cline{3-9}
 &  & \multirow{1}{1cm}{1} & 14.4 & 3.76 & <0.001  & 14.1 & 3.72 & <0.001 \tabularnewline
\cline{3-9}
 &  & \multirow{1}{1cm}{5} & 20.4 & 4.46 & <0.001  & 20.0 & 4.41 & <0.001 \tabularnewline
\hline 
\end{tabular*}
\end{table}

\subsubsection{PET-PEESE}

For Cohen's d the AMSE of the PET-PEESE method is higher for the equal
study variance scenario than the unequal variance scenario for all
simulation combination. The bias is higher for the equal study variance
scenario when $m=10$ and $n_{i}=15$. The AMSE increases and the
bias decreases when the between-study variance increases. The coverage
probability is, in general, nominal or almost nominal for all simulation
settings. These results hold for both cases of homogeneity and heterogeneity
of the treatment effect. The coverage probability, however, is comparable
for different values of $\tau^{2}$.

\noindent Hedges g shows higher values of AMSE for the equal study
variance scenario than the unequal study variance scenario. The bias
of Hedges g decreases in general as the treatment effect heterogeneity
increases. The statistical power is nominal or almost nominal for
all simulation settings. 

\noindent The AMSE for the Cohen's d and Hedges g are comparable.
When $n_{i}=15$ Hedges g shows lower bias than Cohen's d in case
of equal study variance scenario, but this difference reduces or disappears
when $n_{i}=30$. For the unequal study variance scenario no clear
pattern can be observed. Coverage probabilities are comparable for
Cohen's d and the Hedges g for all simulation scenario's.

\noindent The AMSE and bias of the PET-PEESE method under the equal
variance scenario are higher than under the unequal variance scenario.
The statistical power of the method is higher under the equal variance
scenario.

\medskip{}
\begin{table}[H]
\caption{An overview of the AMSE, bias and coverage probabilities for the PET-PEESE
method under equal variance assumption for the different simulation
settings.}

\medskip{}

\begin{tabular*}{1cm}{@{\extracolsep{\fill}}|>{\raggedright}p{1cm}|>{\raggedright}p{1cm}|>{\raggedright}p{1cm}|>{\raggedright}p{1cm}|>{\raggedright}p{1cm}|>{\raggedright}p{1cm}|>{\raggedright}p{1cm}|>{\raggedright}p{1cm}|>{\raggedright}p{1cm}|}
\hline 
\multirow{2}{1cm}{$m$} & \multirow{2}{1cm}{$n_{i}$} & \multirow{2}{1cm}{$\tau^{2}$} & \multicolumn{3}{c|}{Cohen D} & \multicolumn{3}{c|}{Hedges G}\tabularnewline
\cline{4-9}
 &  &  & \multicolumn{1}{l|}{Amse} & \multicolumn{1}{l|}{Bias} & \multicolumn{1}{l|}{COV} & \multicolumn{1}{l|}{Amse} & \multicolumn{1}{l|}{Bias} & \multicolumn{1}{l|}{COV}\tabularnewline
\hline 
\multirow{6}{1cm}{10} & \multirow{3}{1cm}{15} & \multirow{1}{1cm}{0} & 6.51 & -1.67 & 0.895 & 6.79 & -1.68 & 0.892\tabularnewline
\cline{3-9}
 &  & \multirow{1}{1cm}{1} & 6.96 & -1.67 & 0.879 & 7.12 & -1.68 & 0.879\tabularnewline
\cline{3-9}
 &  & \multirow{1}{1cm}{5} & 2.13 & -1.08 & 0.829 & 7.57 & -1.66 & 0.896\tabularnewline
\cline{2-9}
 & \multirow{3}{1cm}{30} & \multirow{1}{1cm}{0} & 2.33 & -0.851 & 0.825 & 2.41 & -0.926 & 0.819\tabularnewline
\cline{3-9}
 &  & \multirow{1}{1cm}{1} & 2.47 & -0.877 & 0.820 & 2.45 & -0.832 & 0.828\tabularnewline
\cline{3-9}
 &  & \multirow{1}{1cm}{5} & 4.61 & -1.16 & 0.850 & 4.43 & -1.26 & 0.849\tabularnewline
\hline 
\multirow{6}{1cm}{30} & \multirow{3}{1cm}{15} & \multirow{1}{1cm}{0} & 2.01 & -1.04 & 0.827 & 2.02 & -1.13 & 0.790\tabularnewline
\cline{3-9}
 &  & \multirow{1}{1cm}{1} & 2.13 & -1.08 & 0.829 & 2.19 & -1.17 & 0.782\tabularnewline
\cline{3-9}
 &  & \multirow{1}{1cm}{5} & 2.45 & -1.24 & 0.804 & 2.39 & -1.24 & 0.806\tabularnewline
\cline{2-9}
 & \multirow{3}{1cm}{30} & \multirow{1}{1cm}{0} & 0.702 & -0.617 & 0.718 & 0.728 & -0.639 & 0.700\tabularnewline
\cline{3-9}
 &  & \multirow{1}{1cm}{1} & 0.715 & -0.624 & 0.725 & 0.662 & -0.567 & 0.744\tabularnewline
\cline{3-9}
 &  & \multirow{1}{1cm}{5} & 0.982 & -0.700 & 0.81200 & 1.045 & -0.775 & 0.786\tabularnewline
\hline 
\end{tabular*}
\end{table}

\medskip{}
\begin{table}[H]
\caption{An overview of the AMSE, bias and coverage probabilities for the PET-PEESE
method under unequal variance assumption for the different simulation
settings.}

\medskip{}

\begin{tabular*}{1cm}{@{\extracolsep{\fill}}|>{\raggedright}p{1cm}|>{\raggedright}p{1cm}|>{\raggedright}p{1cm}|>{\raggedright}p{1cm}|>{\raggedright}p{1cm}|>{\raggedright}p{1cm}|>{\raggedright}p{1cm}|>{\raggedright}p{1cm}|>{\raggedright}p{1cm}|}
\hline 
\multirow{2}{1cm}{$m$} & \multirow{2}{1cm}{$n_{i}$} & \multirow{2}{1cm}{$\tau^{2}$} & \multicolumn{3}{c|}{Cohen D} & \multicolumn{3}{c|}{Hedges G}\tabularnewline
\cline{4-9}
 &  &  & \multicolumn{1}{l|}{Amse} & \multicolumn{1}{l|}{Bias} & \multicolumn{1}{l|}{COV} & \multicolumn{1}{l|}{Amse} & \multicolumn{1}{l|}{Bias} & \multicolumn{1}{l|}{COV}\tabularnewline
\hline 
\multirow{6}{1cm}{10} & \multirow{3}{1cm}{15} & \multirow{1}{1cm}{0} & 5.59 & -1.17 & \multirow{1}{1cm}{0.929} & 5.42 & -1.18 & 0.929\tabularnewline
\cline{3-9}
 &  & \multirow{1}{1cm}{1} & 5.44 & -1.14 & 0.938 & 5.33 & -1.15 & 0.930\tabularnewline
\cline{3-9}
 &  & \multirow{1}{1cm}{5} & 6.89 & -1.26 & 0.923 & 6.73 & -1.28 & 0.925\tabularnewline
\cline{2-9}
 & \multirow{3}{1cm}{30} & \multirow{1}{1cm}{0} & 1.49 & -0.121 & 0.785 & 1.52 & -0.086 & 0.781\tabularnewline
\cline{3-9}
 &  & \multirow{1}{1cm}{1} & 1.50 & -0.110 & 0.779 & 1.53 & -0.080 & 0.775\tabularnewline
\cline{3-9}
 &  & \multirow{1}{1cm}{5} & 2.60 & -0.560 & 0.837 & 2.58 & -0.573 & 0.837\tabularnewline
\hline 
\multirow{6}{1cm}{30} & \multirow{3}{1cm}{15} & \multirow{1}{1cm}{0} & 1.09 & -0.564 & 0.973 & 1.04 & -0.577 & 0.969\tabularnewline
\cline{3-9}
 &  & \multirow{1}{1cm}{1} & 1.12 & -0.568 & 0.967 & 1.10 & -0.589 & 0.965\tabularnewline
\cline{3-9}
 &  & \multirow{1}{1cm}{5} & 1.41 & -0.676 & 0.967 & 1.38 & -0.691 & 0.964\tabularnewline
\cline{2-9}
 & \multirow{3}{1cm}{30} & \multirow{1}{1cm}{0} & 0.346 & 0.091 & 0.809 & 0.340 & 0.069 & 0.816\tabularnewline
\cline{3-9}
 &  & \multirow{1}{1cm}{1} & 0.344 & 0.081 & 0.809 & 0.339 & 0.059 & 0.813\tabularnewline
\cline{3-9}
 &  & \multirow{1}{1cm}{5} & 0.436 & -0.183 & 0.929 & 0.437 & -0.200 & 0.929\tabularnewline
\hline 
\end{tabular*}
\end{table}

\subsubsection{Trim \& Fill Method}

For the Cohen's d the AMSE and bias of the Trim \& Fill method are
higher for the equal study variance scenario when $n_{i}=15$ but
not when $n_{i}=30$. The AMSE and bias increase when $\tau^{2}=5$
for the equal and unequal study variances scenario's. The coverage
probability for Cohen's d is comparable for the equal and unequal
study variance scenario's. Hedges g shows higher values of AMSE and
bias and when $m=10$ and $n_{i}=15$ for the equal study variance
scenario for different values of $\tau^{2}$ than when $m=30$. The
AMSE and bias increase when $\tau^{2}=5$ for the equal and unequal
study variances scenario's. The AMSE and bias are generally higher
for the equal study variances scenario than the unequal study variances
scenario. The coverage probabilities for Hedges g are comparable for
the equal and unequal study variance scenario's. However, the coverage
probability decreases when $m=30$ and $n_{i}=30$. The AMSE and bias
of Cohen's d are higher that those of Hedges g for the equal study
variances scenario except when $m=30$ and $n_{i}=30$ and also when
$\tau^{2}=5$. The coverage probabilities of Cohen's d and Hedges
g have comparable values except when $\tau^{2}=5$ for the equal study
variance scenario where Hedges g has a higher coverage probability.

\noindent The AMSE and bias of the Trim \& Fill method under the equal
study variance scenario are lower than under the unequal variance
scenario. The coverage probability of the method is lower under the
equal variance scenario.

\medskip{}
\begin{table}[H]
\caption{An overview of the AMSE, bias and coverage probabilities for the Trim
\& Fill method under equal variance assumption for the different simulation
settings.}

\medskip{}

\begin{tabular*}{1cm}{@{\extracolsep{\fill}}|>{\raggedright}p{1cm}|>{\raggedright}p{1cm}|>{\raggedright}p{1cm}|>{\raggedright}p{1cm}|>{\raggedright}p{1cm}|>{\raggedright}p{1cm}|>{\raggedright}p{1cm}|>{\raggedright}p{1cm}|>{\raggedright}p{1cm}|}
\hline 
\multirow{2}{1cm}{$m$} & \multirow{2}{1cm}{$n_{i}$} & \multirow{2}{1cm}{$\tau^{2}$} & \multicolumn{3}{c|}{Cohen D} & \multicolumn{3}{c|}{Hedges G}\tabularnewline
\cline{4-9}
 &  &  & \multicolumn{1}{l|}{Amse} & \multicolumn{1}{l|}{Bias} & \multicolumn{1}{l|}{COV} & \multicolumn{1}{l|}{Amse} & \multicolumn{1}{l|}{Bias} & \multicolumn{1}{l|}{COV}\tabularnewline
\hline 
\multirow{6}{1cm}{10} & \multirow{3}{1cm}{15} & \multirow{1}{1cm}{0} & 5.00 & 1.85 & 0.530 & 4.98 & 1.86 & 0.521\tabularnewline
\cline{3-9}
 &  & \multirow{1}{1cm}{1} & 4.98 & 1.85 & 0.531 & 4.97 & 1.86 & 0.521\tabularnewline
\cline{3-9}
 &  & \multirow{1}{1cm}{5} & 3.34 & 1.65 & 0.271 & 5.61 & 1.93 & 0.545\tabularnewline
\cline{2-9}
 & \multirow{3}{1cm}{30} & \multirow{1}{1cm}{0} & 8.14 & 2.71 & 0.116 & 7.64 & 2.62 & 0.129\tabularnewline
\cline{3-9}
 &  & \multirow{1}{1cm}{1} & 8.08 & 2.69 & 0.125 & 8.14 & 2.71 & 0.123\tabularnewline
\cline{3-9}
 &  & \multirow{1}{1cm}{5} & 10.5 & 2.97 & 0.230 & 9.42 & 2.79 & 0.266\tabularnewline
\hline 
\multirow{6}{1cm}{30} & \multirow{3}{1cm}{15} & \multirow{1}{1cm}{0} & 3.36 & 1.66 & 0.264 & 2.67 & 1.46 & 0.322\tabularnewline
\cline{3-9}
 &  & \multirow{1}{1cm}{1} & 3.34 & 1.65 & 0.271 & 2.64 & 1.45 & 0.326\tabularnewline
\cline{3-9}
 &  & \multirow{1}{1cm}{5} & 2.98 & 1.52 & 0.369 & 2.97 & 1.52 & 0.359\tabularnewline
\cline{2-9}
 & \multirow{3}{1cm}{30} & \multirow{1}{1cm}{0} & 6.67 & 2.50 & 0.011 & 6.47 & 2.46 & 0.010\tabularnewline
\cline{3-9}
 &  & \multirow{1}{1cm}{1} & 6.72 & 2.51 & 0.006 & 6.98 & 2.56 & 0.007\tabularnewline
\cline{3-9}
 &  & \multirow{1}{1cm}{5} & 8.11 & 2.69 & 0.051 & 7.42 & 2.57 & 0.057\tabularnewline
\hline 
\end{tabular*}
\end{table}

\medskip{}
\begin{table}[H]
\caption{An overview of the AMSE, bias and coverage probabilities for the Trim
\& Fill method under unequal variance assumption for the different
simulation settings.}

\medskip{}

\begin{tabular*}{1cm}{@{\extracolsep{\fill}}|>{\raggedright}p{1cm}|>{\raggedright}p{1cm}|>{\raggedright}p{1cm}|>{\raggedright}p{1cm}|>{\raggedright}p{1cm}|>{\raggedright}p{1cm}|>{\raggedright}p{1cm}|>{\raggedright}p{1cm}|>{\raggedright}p{1cm}|}
\hline 
\multirow{2}{1cm}{$m$} & \multirow{2}{1cm}{$n_{i}$} & \multirow{2}{1cm}{$\tau^{2}$} & \multicolumn{3}{c|}{Cohen D} & \multicolumn{3}{c|}{Hedges G}\tabularnewline
\cline{4-9}
 &  &  & \multicolumn{1}{l|}{Amse} & \multicolumn{1}{l|}{Bias} & \multicolumn{1}{l|}{COV} & \multicolumn{1}{l|}{Amse} & \multicolumn{1}{l|}{Bias} & \multicolumn{1}{l|}{COV}\tabularnewline
\hline 
\multirow{6}{1cm}{10} & \multirow{3}{1cm}{15} & \multirow{1}{1cm}{0} & 7.49 & 2.45 & \multirow{1}{1cm}{0.344} & 7.18 & 2.40 & 0.344\tabularnewline
\cline{3-9}
 &  & \multirow{1}{1cm}{1} & 7.57 & 2.47 & 0.345 & 7.25 & 2.41 & 0.346\tabularnewline
\cline{3-9}
 &  & \multirow{1}{1cm}{5} & 8.68 & 2.60 & 0.375 & 8.29 & 2.54 & 0.380\tabularnewline
\cline{2-9}
 & \multirow{3}{1cm}{30} & \multirow{1}{1cm}{0} & 12.1 & 3.38 & 0.025 & 12.1 & 3.39 & 0.025\tabularnewline
\cline{3-9}
 &  & \multirow{1}{1cm}{1} & 12.3 & 3.39 & 0.027 & 12.2 & 3.39 & 0.030\tabularnewline
\cline{3-9}
 &  & \multirow{1}{1cm}{5} & 14.1 & 3.53 & 0.131 & 13.8 & 3.49 & 0.132\tabularnewline
\hline 
\multirow{6}{1cm}{30} & \multirow{3}{1cm}{15} & \multirow{1}{1cm}{0} & 4.82 & 2.07 & 0.082 & 4.63 & 2.03 & 0.080\tabularnewline
\cline{3-9}
 &  & \multirow{1}{1cm}{1} & 4.86 & 2.08 & 0.084 & 4.61 & 2.02 & 0.087\tabularnewline
\cline{3-9}
 &  & \multirow{1}{1cm}{5} & 5.13 & 2.11 & 0.123 & 4.94 & 2.07 & 0.129\tabularnewline
\cline{2-9}
 & \multirow{3}{1cm}{30} & \multirow{1}{1cm}{0} & 10.8 & 3.23 & <0.001  & 10.5 & 3.19 & <0.001 \tabularnewline
\cline{3-9}
 &  & \multirow{1}{1cm}{1} & 10.9 & 3.23 & <0.001  & 10.6 & 3.19 & <0.001 \tabularnewline
\cline{3-9}
 &  & \multirow{1}{1cm}{5} & 11.4 & 3.25 & 0.011 & 11.2 & 3.21 & 0.010\tabularnewline
\hline 
\end{tabular*}
\end{table}

\subsubsection{The limit meta-analysis method}

For Cohen's d the AMSE and bias of the limit meta-analysis method
are higher for the equal study variance scenario except when $m=30$
and $n_{i}=30$. The bias increases as the treatment effect heterogeneity
increases, especially when $\tau^{2}=5$. The coverage probability
for Cohen's d is comparable for the equal and unequal study variance
scenario's.\textbf{ }For the equal variance scenario, the coverage
probability in general is above the nominal level, with the exception
of the case $m=10$ and $n_{i}=30$ where it becomes very low. In
case of the unequal variance scenario when $n=10$, the coverage probability
is above the nominal level except for the cases $n_{i}=15$ and $\tau^{2}=5$,
and $n_{i}=30$ and $\tau^{2}=1$. When $m=30$ the coverage probability
is either equal or slightly less than the nominal value. For the equal
variance scenario the AMSE and bias of the Cohen's d are higher than
those of the Hedges g when $n_{i}=15$. The values of AMSE and bias
of both methods are comparable for the unequal variance scenario.
Hedges g shows higher values of AMSE and bias when $n_{i}=15$ for
both the equal and unequal study variance scenario's. For both methods,
the AMSE and bias increase as the treatment effect heterogeneity increases
for both the equal and unequal study variances scenario's. The coverage
probability for Hedges g is comparable for the equal and unequal study
variance scenario's, but it is slightly liberal when $m=10$ and slightly
conservative when $m=30$. Coverage probabilities are comparable for
the Cohen's d and the Hedges g methods for all simulation scenario's.

\noindent The AMSE and bias of the limit meta-analysis method under
the equal variance scenario are lower than those under the unequal
variance scenario. The statistical power of the method under the equal
and unequal variance scenario's are comparable. 

\medskip{}
\begin{table}[H]
\caption{An overview of the AMSE, bias and coverage probabilities for the limit
meta-analysis method under equal variance assumption for the different
simulation settings.}

\medskip{}

\begin{tabular*}{1cm}{@{\extracolsep{\fill}}|>{\raggedright}p{1cm}|>{\raggedright}p{1cm}|>{\raggedright}p{1cm}|>{\raggedright}p{1cm}|>{\raggedright}p{1cm}|>{\raggedright}p{1cm}|>{\raggedright}p{1cm}|>{\raggedright}p{1cm}|>{\raggedright}p{1cm}|}
\hline 
\multirow{2}{1cm}{$m$} & \multirow{2}{1cm}{$n_{i}$} & \multirow{2}{1cm}{$\tau^{2}$} & \multicolumn{3}{c|}{Cohen D} & \multicolumn{3}{c|}{Hedges G}\tabularnewline
\cline{4-9}
 &  &  & \multicolumn{1}{l|}{Amse} & \multicolumn{1}{l|}{Bias} & \multicolumn{1}{l|}{COV} & \multicolumn{1}{l|}{Amse} & \multicolumn{1}{l|}{Bias} & \multicolumn{1}{l|}{COV}\tabularnewline
\hline 
\multirow{6}{1cm}{10} & \multirow{3}{1cm}{15} & \multirow{1}{1cm}{0} & 3.32 & 1.67 & 0.992 & 3.25 & 1.64 & 0.992\tabularnewline
\cline{3-9}
 &  & \multirow{1}{1cm}{1} & 3.37 & 1.68 & 0.995 & 3.28 & 1.65 & 0.994\tabularnewline
\cline{3-9}
 &  & \multirow{1}{1cm}{5} & 3.95 & 1.84 & 0.914 & 3.89 & 1.81 & 0.912\tabularnewline
\cline{2-9}
 & \multirow{3}{1cm}{30} & \multirow{1}{1cm}{0} & 1.70 & 1.18 & 0.503 & 1.62 & 1.16 & 0.503\tabularnewline
\cline{3-9}
 &  & \multirow{1}{1cm}{1} & 1.67 & 1.17 & 1.000 & 1.64 & 1.15 & 1.000\tabularnewline
\cline{3-9}
 &  & \multirow{1}{1cm}{5} & 3.62 & 1.66  & 0.999 & 3.40 & 1.61 & 0.999\tabularnewline
\hline 
\multirow{6}{1cm}{30} & \multirow{3}{1cm}{15} & \multirow{1}{1cm}{0} & 0.464 & 0.646 & 0.965 & 0.392 & 0.593 & 0.972\tabularnewline
\cline{3-9}
 &  & \multirow{1}{1cm}{1} & 0.470 & 0.652 & 0.965 & 0.396 & 0.598 & 0.971\tabularnewline
\cline{3-9}
 &  & \multirow{1}{1cm}{5} & 0.575 & 0.729 & 0.977 & 0.539 & 0.703 & 0.973\tabularnewline
\cline{2-9}
 & \multirow{3}{1cm}{30} & \multirow{1}{1cm}{0} & 0.157 & 0.368 & 0.995 & 0.150 & 0.359 & 0.995\tabularnewline
\cline{3-9}
 &  & \multirow{1}{1cm}{1} & 0.165 & 0.379 & 0.992 & 0.160 & 0.371 & 0.991\tabularnewline
\cline{3-9}
 &  & \multirow{1}{1cm}{5} & 0.445 & 0.637 & 0.990 & 0.417 & 0.617 & 0.991\tabularnewline
\hline 
\end{tabular*}
\end{table}

\medskip{}
\begin{table}[H]
\caption{An overview of the AMSE, bias and coverage probabilities for the limit
meta-analysis method under unequal variance assumption for the different
simulation settings.}

\medskip{}

\begin{tabular*}{1cm}{@{\extracolsep{\fill}}|>{\raggedright}p{1cm}|>{\raggedright}p{1cm}|>{\raggedright}p{1cm}|>{\raggedright}p{1cm}|>{\raggedright}p{1cm}|>{\raggedright}p{1cm}|>{\raggedright}p{1cm}|>{\raggedright}p{1cm}|>{\raggedright}p{1cm}|}
\hline 
\multirow{2}{1cm}{$m$} & \multirow{2}{1cm}{$n_{i}$} & \multirow{2}{1cm}{$\tau^{2}$} & \multicolumn{3}{c|}{Cohen D} & \multicolumn{3}{c|}{Hedges G}\tabularnewline
\cline{4-9}
 &  &  & \multicolumn{1}{l|}{AMSE} & \multicolumn{1}{l|}{BIAS} & \multicolumn{1}{l|}{COV} & \multicolumn{1}{l|}{AMSE} & \multicolumn{1}{l|}{BIAS} & \multicolumn{1}{l|}{COV}\tabularnewline
\hline 
\multirow{6}{1cm}{10} & \multirow{3}{1cm}{15} & \multirow{1}{1cm}{0} & 5.06 & 2.13 & \multirow{1}{1cm}{0.982} & 4.78 & 2.07 & 0.981\tabularnewline
\cline{3-9}
 &  & \multirow{1}{1cm}{1} & 5.08 & 2.13 & 0.982 & 4.82 & 2.07 & 0.981\tabularnewline
\cline{3-9}
 &  & \multirow{1}{1cm}{5} & 6.23 & 2.35 & 0.459 & 5.92 & 2.30 & 0.459\tabularnewline
\cline{2-9}
 & \multirow{3}{1cm}{30} & \multirow{1}{1cm}{0} & 2.73 & 1.58 & 0.995 & 2.75 & 1.57 & 0.992\tabularnewline
\cline{3-9}
 &  & \multirow{1}{1cm}{1} & 2.77 & 1.59 & 0.990 & 2.84 & 1.58 & 0.692\tabularnewline
\cline{3-9}
 &  & \multirow{1}{1cm}{5} & 4.76 & 2.08  & 0.355 & 4.69 & 2.05 & 0.992\tabularnewline
\hline 
\multirow{6}{1cm}{30} & \multirow{3}{1cm}{15} & \multirow{1}{1cm}{0} & 1.22 & 1.09 & 0.950 & 1.16 & 1.06 & 0.948\tabularnewline
\cline{3-9}
 &  & \multirow{1}{1cm}{1} & 1.23 & 1.09 & 0.949 & 1.16 & 1.06 & 0.948\tabularnewline
\cline{3-9}
 &  & \multirow{1}{1cm}{5} & 1.52 & 1.21 & 0.956 & 1.44 & 1.18 & 0.955\tabularnewline
\cline{2-9}
 & \multirow{3}{1cm}{30} & \multirow{1}{1cm}{0} & 0.716 & 0.834 & 0.920 & 0.700 & 0.825 & 0.920\tabularnewline
\cline{3-9}
 &  & \multirow{1}{1cm}{1} & 0.740 & 0.848 & 0.918 & 0.724 & 0.838 & 0.917\tabularnewline
\cline{3-9}
 &  & \multirow{1}{1cm}{5} & 1.29 & 1.12 & 0.978 & 1.261 & 1.12 & 0.977\tabularnewline
\hline 
\end{tabular*}
\end{table}

\subsubsection{Random effects model using the DerSimonian-Laird estimate}

For Cohen's d the AMSE and bias of the random effects model are higher
for the equal study variance scenario when $n_{i}=15$ but not when
$n_{i}=30$. The AMSE and bias increase when $\tau^{2}=5$ for the
equal and unequal study variances scenario's. The coverage probability
for Cohen's d is comparable for the equal and unequal study variance
scenario's. Hedges g shows lower values of AMSE and bias and when
$n_{i}=15$ for the equal study variance scenario for different values
of $\tau^{2}$ than when $n_{i}=30$ for both the equal and unequal
variance scenario's. The AMSE and bias of Hedges g increase when $\tau^{2}=5$
for both the equal and unequal study variances scenario's. The AMSE
and bias are generally higher for the unequal study variances scenario
than for the equal study variances scenario. The coverage probabilities
for both Cohen's d and Hedges g are generally low and they decrease
when $n_{i}=30$. The coverage probabilities of Cohen's d and Hedges
g have comparable values except when $\tau^{2}=5$ for the equal study
variance scenario where Hedges g has a higher coverage probability.

\noindent The AMSE and bias of the random effects model using the
DerSimonian-Laird estimate method under the equal variance scenario
are lower than those under the unequal variance scenario. The statistical
power of the method is lower under the equal variance scenario.

\medskip{}
\begin{table}[H]
\caption{An overview of the AMSE, bias and coverage probabilities for the random
effects model under equal variance assumption for the different simulation
settings.}

\medskip{}

\begin{tabular*}{1cm}{@{\extracolsep{\fill}}|>{\raggedright}p{1cm}|>{\raggedright}p{1cm}|>{\raggedright}p{1cm}|>{\raggedright}p{1cm}|>{\raggedright}p{1cm}|>{\raggedright}p{1cm}|>{\raggedright}p{1cm}|>{\raggedright}p{1cm}|>{\raggedright}p{1cm}|}
\hline 
\multirow{2}{1cm}{$m$} & \multirow{2}{1cm}{$n_{i}$} & \multirow{2}{1cm}{$\tau^{2}$} & \multicolumn{3}{c|}{Cohen D} & \multicolumn{3}{c|}{Hedges G}\tabularnewline
\cline{4-9}
 &  &  & \multicolumn{1}{l|}{Amse} & \multicolumn{1}{l|}{Bias} & \multicolumn{1}{l|}{COV} & \multicolumn{1}{l|}{Amse} & \multicolumn{1}{l|}{Bias} & \multicolumn{1}{l|}{COV}\tabularnewline
\hline 
\multirow{6}{1cm}{10} & \multirow{3}{1cm}{15} & \multirow{1}{1cm}{0} & 6.10 & 2.11 & 0.574 & 6.00 & 2.11 & 0.559\tabularnewline
\cline{3-9}
 &  & \multirow{1}{1cm}{1} & 6.14 & 2.11 & 0.579 & 6.05 & 2.12 & 0.561\tabularnewline
\cline{3-9}
 &  & \multirow{1}{1cm}{5} & 6.81 & 2.18 & 0.617 & 6.74 & 2.18 & 0.598\tabularnewline
\cline{2-9}
 & \multirow{3}{1cm}{30} & \multirow{1}{1cm}{0} & 8.98 & 2.88 & 0.096 & 8.48 & 2.80 & 0.109\tabularnewline
\cline{3-9}
 &  & \multirow{1}{1cm}{1} & 9.02 & 2.88 & 0.110 & 9.08 & 2.90 & 0.099\tabularnewline
\cline{3-9}
 &  & \multirow{1}{1cm}{5} & 11.8 & 3.22 & 0.268 & 10.6 & 3.02 & 0.305\tabularnewline
\hline 
\multirow{6}{1cm}{30} & \multirow{3}{1cm}{15} & \multirow{1}{1cm}{0} & 5.93 & 2.31 & 0.154 & 4.69 & 2.03 & 0.221\tabularnewline
\cline{3-9}
 &  & \multirow{1}{1cm}{1} & 5.94 & 2.32 & 0.161 & 4.70 & 2.04 & 0.225\tabularnewline
\cline{3-9}
 &  & \multirow{1}{1cm}{5} & 5.24 & 2.13 & 0.308 & 5.14 & 2.11 & 0.28100\tabularnewline
\cline{2-9}
 & \multirow{3}{1cm}{30} & \multirow{1}{1cm}{0} & 8.37 & 2.85 & <0.001  & 8.14 & 2.81 & <0.001 \tabularnewline
\cline{3-9}
 &  & \multirow{1}{1cm}{1} & 8.44 & 2.86 & <0.001  & 8.62 & 2.89 & <0.001 \tabularnewline
\cline{3-9}
 &  & \multirow{1}{1cm}{5} & 10.8 & 3.20 & 0.015 & 9.85 & 3.05 & 0.018\tabularnewline
\hline 
\end{tabular*}
\end{table}

\medskip{}
\begin{table}[H]
\caption{An overview of the AMSE, bias and coverage probabilities for the random
effects model under unequal variance assumption for the different
simulation settings.}

\medskip{}

\begin{tabular*}{1cm}{@{\extracolsep{\fill}}|>{\raggedright}p{1cm}|>{\raggedright}p{1cm}|>{\raggedright}p{1cm}|>{\raggedright}p{1cm}|>{\raggedright}p{1cm}|>{\raggedright}p{1cm}|>{\raggedright}p{1cm}|>{\raggedright}p{1cm}|>{\raggedright}p{1cm}|}
\hline 
\multirow{2}{1cm}{$m$} & \multirow{2}{1cm}{$n_{i}$} & \multirow{2}{1cm}{$\tau^{2}$} & \multicolumn{3}{c|}{Cohen D} & \multicolumn{3}{c|}{Hedges G}\tabularnewline
\cline{4-9}
 &  &  & \multicolumn{1}{l|}{AMSE} & \multicolumn{1}{l|}{BIAS} & \multicolumn{1}{l|}{COV} & \multicolumn{1}{l|}{AMSE} & \multicolumn{1}{l|}{BIAS} & \multicolumn{1}{l|}{COV}\tabularnewline
\hline 
\multirow{6}{1cm}{10} & \multirow{3}{1cm}{15} & \multirow{1}{1cm}{0} & 9.04 & 2.75 & \multirow{1}{1cm}{0.401} & 8.51 & 2.67 & 0.403\tabularnewline
\cline{3-9}
 &  & \multirow{1}{1cm}{1} & 9.14 & 2.76 & 0.397 & 8.62 & 2.68 & 0.396\tabularnewline
\cline{3-9}
 &  & \multirow{1}{1cm}{5} & 10.4 & 2.88 & 0.448 & 9.80 & 2.80 & 0.451\tabularnewline
\cline{2-9}
 & \multirow{3}{1cm}{30} & \multirow{1}{1cm}{0} & 13.1 & 3.54 & 0.029 & 13.1 & 3.53 & 0.027\tabularnewline
\cline{3-9}
 &  & \multirow{1}{1cm}{1} & 13.3 & 3.55 & 0.026 & 13.2 & 3.55167 & 0.026\tabularnewline
\cline{3-9}
 &  & \multirow{1}{1cm}{5} & 15.8 & 3.77 & 0.166 & 15.4 & 3.72 & 0.165\tabularnewline
\hline 
\multirow{6}{1cm}{30} & \multirow{3}{1cm}{15} & \multirow{1}{1cm}{0} & 7.78 & 2.69 & 0.062 & 7.33 & 2.61 & 0.062\tabularnewline
\cline{3-9}
 &  & \multirow{1}{1cm}{1} & 7.78 & 2.69 & 0.059 & 7.34 & 2.61 & 0.058\tabularnewline
\cline{3-9}
 &  & \multirow{1}{1cm}{5} & 8.47 & 2.78 & 0.101 & 8.00 & 2.71 & 0.100\tabularnewline
\cline{2-9}
 & \multirow{3}{1cm}{30} & \multirow{1}{1cm}{0} & 12.7 & 3.54 & <0.001  & 12.4 & 3.49 & <0.001 \tabularnewline
\cline{3-9}
 &  & \multirow{1}{1cm}{1} & 12.9 & 3.55 & <0.001  & 12.6 & 3.51 & <0.001 \tabularnewline
\cline{3-9}
 &  & \multirow{1}{1cm}{5} & 14.4 & 3.73 & <0.001  & 14.1 & 3.68 & <0.001 \tabularnewline
\hline 
\end{tabular*}
\end{table}

\subsection{Comparison of results of the methods correcting for publication bias}

\subsubsection{AMSE}

The p uniform has the lowest AMSE followed by the Copas the Trim \&
Fill methods. The highest AMSE was found for the PET-PEESE and the
limit meta-analysis methods, respectively. 

\subsubsection{Bias}

The PET PEESE is negatively biased while the limit meta-analysis method
is mostly positively biased except in the case of $m=30$. While generally
negatively biased when $m=30$, the limit meta analysis method has
lower bias than the PET PEESE. The Copas method is positively biased
but has a relatively small bias. The Trim \& Fill and the random-effects
meta-analysis methods are positively biased and have comparable biases.
The p-uniform method is also positively biased but has the biggest
bias among all the methods.

\subsubsection{Coverage probability}

The PET PEESE has coverage probability usually below the nominal value,
except in the case of $m=10$ and $n_{i}=15$. As mentioned above,
the limit meta-analysis has a liberal coverage probability when $m=10$
but a slightly conservative coverage probability when $m=30$. The
Copas and the Trim \& Fill methods have comparable coverage probability.
Both methods have clearly a lower than nominal coverage probability,
and their coverage probabilities decrease when $n_{i}$=30. The p
uniform method and the random-effects meta-analysis methods have the
lowest coverage probability.

\section{Discussion}

\noindent The purpose of this article was to compare the performance
of five methods adjusting for publication bias. The five methods were
Copas, p-uniform, PET-PEESE, Trim \& Fill, and limit meta methods.
In addition, the random effect meta-analysis method based on the DerSimonian-Laird
estimate was also included. This study focused on continuous outcomes
and investigated the adjustment methods using two treatment effect
measures: Cohen's d and Hedges g. The performance of the adjustment
methods were compared using a case-study and a simulation study. The
simulation settings included different scenario's, including different
number of studies, different number of individuals per study, treatment
effect homogeneity and different levels of treatment effect heterogeneity.
In addition, the two scenario's of equal and unequal variances for
the treatment and control groups were also investigated. The performance
of the adjustment methods was investigated by calculating the average
mean squared error, the bias and the coverage probability. 

\noindent The PET-PEESE method performs slightly worse as the treatment
effect heterogeneity increases, a result stated earlier in the literature
\cite{key-1-6}. The p-uniform method overestimates the treatment
effect and its performance deteriorates as the treatment effect heterogeneity
increases, a result presented earlier in the literature \cite{key-1-1-8}.
The limit meta analysis is too conservative compared to the other
methods investigated. This probably is the result of the equivalence
of the method to a regression analysis and that the limits are basically
prediction limits. These prediction limits not only take the precision
of the mean value into account, but also the precision of the between-study
variance component and that of the parameter representing publication
bias. 

\noindent Copas\textquoteright{} method has been compared to the Trim
and Fill method using the odds ratio. The Copas\textquoteright{} method
was preferred since it produced smaller standard errors \cite{key-1-1-1-1-12}.
Our analysis for Cohen's d and Hedges g also shows that the Copas
method has smaller bias and higher coverage probabilities than the
Trim \& Fill method. However, the Copas method performs less well
as treatment effect heterogeneity increases. 

\noindent In general, no clear difference in performance was observed
between Cohen's d and Hedges g for all methods in both cases of equal
and unequal variance scenario's. With the exception of the PET-PEESE
method, both Cohen's d and Hedges g perform better under the equal
variance scenario for all methods. This could be explained by the
fact that both methods assume equal variances and use a pooled variance.
PET-PEESE method's better performance in the unequal variance scenario
could be explained by the fact that both of the method's estimates
are based on weighted regression models, which take heteroscedasticty
into account.

\noindent Disregarding the direction of the estimation bias the Copas
and PET-PEESE methods are the least biased methods. However, the PET-PEESE
was the only method showing no treatment effect in the case-study
analyzed, in agreement with the conclusion that this specific meta-analysis
suffered from publication bias \cite{key-1-2-1-2}. In addition, the
Copas method, like all Likelihood-based methods, can have convergence
problems, which gives the PET-PEESE an advantage. 

\part*{Conflict of interest}

The author has declared no conflict of interest.

\newpage{}


\begin{thebibliography}{10}
\bibitem{key-1-1-1-1-1-1} Hedges L.V. Modeling publication selection
effects in meta-analysis. Statistical Science 1992; 7:246 \textendash 255.
Duval S. \& Tweedie R. A nonparametric \textquotedbl Trim and Fill\textquotedbl{}
method of accounting for publication bias in meta-analysis, \textit{Journal
of the American Statistical Association} 2000a; \textbf{95}:89\textendash 98

\bibitem{key-1-1-1-1-1-2} Duval S. \& Tweedie R. Trim and Fill: A
simple funnel-plot-based method of testing and adjusting for publication
bias in meta-analysis, \textit{Biometrics} 2000b; \textbf{56}:455\textendash 63.
Hedges L.V. Modeling publication selection effects in meta-analysis.
\textit{Stat Sci} 1992; 7:246 \textendash 255.

\bibitem{key-1-1-1-1-1-3} Duval S. \& Tweedie R. Trim and Fill: A
simple funnel-plot-based method of testing and adjusting for publication
bias in meta-analysis, \textit{Biometrics} 2000b; \textbf{56}:455\textendash 63.

\bibitem{key-1-1-1-1-1-4} Terrin N., Schmid C.H., Lau J. \& Olkin
I. Adjusting for publication bias in the presence of heterogeneity.
\textit{Statist. Med} 2003; 22:2113\textendash 2126.

\bibitem{key-1-1-1-1-1-5} Peters J.L., Sutton A.J., Jones D.R., Abrams
K.R. \& Rushton L. Performance of the trim and fill method in the
presence of publication bias and between-study heterogeneity. \textit{Statist
Med} 2007, 10, 4544\textendash 4562. DOI: 10.1002/sim.2889.

\bibitem{key-1-1-1-1-1} Iyengar S., \& Greenhouse, J. B. Selection
models and the file drawer problem. \textit{Stat Sci} 1988; 3: 109\textendash 117.

\bibitem{key-1-1-1-1-2-1} Hedges L.V. Distribution theory for Glass\textquoteright s
estimator of effect size and related estimators. \textit{Journal of
Educational Statistics} 1981;6:107-128.

\bibitem{key-1-1-1-1-1-3-1} Borenstein M.. Effect sizes for continuous
data. In Cooper H, Hedges L. V. and Valentine J. C. Effect sizes for
dichtomous data. In The Handbook of Research Synthesis and Meta-Analysis
2nd ed. 2008;221-235.

\bibitem{key-1-1-1-1-2} Vevea, J.L., \& Hedges, L.V. A general linear
model for estimating effect size in the presence of publication bias.
\textit{Psychometrika} 1995; 60: 419\textendash 435.

\bibitem{key-1-1-1-1-3-1} Vevea J.L. \& Woods C.M. Publication Bias
in Research Synthesis: Sensitivity Analysis Using A Priori Weight
Functions. \textit{Psychol Methods} 2005:10(4):428\textendash 443.

\bibitem{key-1-1-1-1-3-2} Dear, K.B., \& Begg B. An approach for
assessing publication bias prior to performing a meta-analysis. \textit{Stat
Sci} 1992;7:237\textendash 245.

\bibitem{key-1-1-1-1-4} Copas J. \& Shi J.Q. Meta-analysis, funnel
plots and sensitivity analysis. \textit{Biostatistics} 2000a,;1:247\textendash 62.
DOI: 10.1093/biostatistics/1.3.247.

\bibitem{key-1-1-1-1-5} Copas J. \& Shi J.Q. A sensitivity analysis
for publication bias in systematic reviews. \textit{Stat Methods Med
Res} 2000b;10:251\textendash 65. DOI: 10.1177/096228020101000402.

\bibitem{key-1-1-1-1-6} Simonsohn U., Nelson L.D. \& Simmons, J.
P. p-curve and effect size: Correcting for publication bias using
only significant results. Perspectives on Psychological Science 2014;9:666\textendash 681.

\bibitem{key-1-1-1-1-7} van Assen M. A., van Aert R. \& Wicherts,
J. M. Metaanalysis using effect size distributions of only statistically
significant studies. \textit{Psychological Methods} 2015;20:293\textendash 309.

\bibitem{key-1-1-1-1-8} Hedges L.V., \& Vevea, J.L. (2005). Selection
method approaches. In H. R. Rothstein, A. J. Sutton, \& M. Borenstein
(Eds.), Publication bias in meta-analysis: Prevention, assessment
and adjustments (pp. 145\textendash 174). Chichester, England: John
Wiley \& Sons.

\bibitem{key-1-1-1-1-9} Jin, Z.C., Zhou, X.-H., \& He, J. Statistical
methods for dealing with publication bias in meta-analysis. \textit{Stat
Med} 2015;34:343\textendash 360.

\bibitem{key-1-1-1-1-10} McShane B.B., Böckenholt U. \& Hansen K.T.
Adjusting for Publication Bias in Meta-Analysis: An Evaluation of
Selection Methods and Some Cautionary Notes. Perspectives on Psychological
Science 2016;11(5):730\textendash 749.

\bibitem{key-1-1-1-1-10-1} Sutton A.J., Song F., Gilbody S.M. \&
Abrams K.R. Modelling publication bias in meta-analysis: a review.
\textit{Stat Methods Med Res} 2000; 9: 421\textendash 445

\bibitem{key-1-1-1-1-12} Schwarzer G., Carpentera J. \& Rücker G.
Empirical evaluation suggests Copas selection model preferable to
trim-and-fill method for selection bias in meta-analysis. \textit{Journal
of Clinical Epidemiology} (2010);63:282-288.

\bibitem{key-1-1-7} Viechtbauer W. Conducting Meta-Analyses in R
with the metafor Package, \textit{Journal of Statistical Software}
2010; \textbf{36}(3):1-48.

\bibitem{key-1} Sterne J.A., Sutton A.J., Ioannidis J.P. et al. Recommendations
for examining and interpreting funnel plot asymmetry in meta-analyses
of randomised controlled trials. \textit{BMJ} 2011:343:d4002.

\bibitem{key-1-2} Simonsohn U., Nelson L.D. \& Simmons J.P. P-curve:
A Key to The File Drawer. \textit{J Exp Psychol Gen.} 2014;143(2):534-547.
http://dx.doi.org/10.1037/a0033242

\bibitem{key-1-3} Moreno S.G., Sutton A.J., Ades A.E., Cooper N.J.
\& Abrams K.R. Adjusting for publication biases across similar interventions
performed well when compared with gold standard data. \textit{J Clin
Epidemiol.} 2011;64(11):1230-41.

\bibitem{key-1-1-5} Stanley T.D. Meta-Regression Methods for Detecting
and Estimating Empirical Effects in the Presence of Publication Selection.
\textit{Oxford B Econ Stat.} 2008;70(1).

\bibitem{key-1-1-6} Stanley T.D. \& Doucouliagos H. Meta-regression
approximations to reduce publication selection bias. \textit{Res Syn
Meth.} 2013;5(1),60-78.

\bibitem{key-1-6} Alinaghi N. \& Reed W.R. Meta-analysis and publication
bias: How well does the FAT-PET-PEESE procedure work? \textit{Res
Syn Meth} 2018;9(2):285-311.

\bibitem{key-1-1-1-9} Rücker G., Schwarzer G., Carpenter J.R. Binder
H. and Schumacher M. Treatment-effect estimates adjusted for small-study
effects via a limit meta-analysis. \textit{Biostatistics} 2011; 12(1):122-42. 

\bibitem{key-3} Simonsohn U., Nelson L.D. \& Simmons J.P. p-Curve
and Effect Size: Correcting for Publication Bias Using Only Significant
Results. \textit{Perspect Psychol Sci }2014;9(6):666-81.

\bibitem{key-1-1-8} van Assen M.A., van Aert, R. \& Wicherts J.M.
Meta-analysis using effect size distributions of only statistically
significant studies. \textit{Psychol Methods} 2015, 20(3), 293.

\bibitem{key-1-1-9} van Aert R.C.M. Meta-Analysis Methods Correcting
for Publication Bias. 2021, DOI: https://github.com/RobbievanAert/puniform.

\bibitem{key-1-1-20} Reed W.R. A Monte Carlo Analysis of Alternative
Meta- Analysis Estimators in the Presence of Publication Bias. Economics:
The Open-Access, Open-Assessment E-Journal. 2015; 9:(2015-30): 1\textemdash 40.

\bibitem{key-1-1-21} McShane B. B., Böckenholt U. \& Hansen K.T.
Adjusting for Publication Bias in Meta-Analysis: An Evaluation of
Selection Methods and Some Cautionary Notes.Perspectives on Psychological
Science. 2016; 11(5):730\textendash{} 749.

\bibitem{key-1-1-1-9-1} Schwarzer G., Carpenter J.R. \& Rucker G.
metasens: Statistical Methods for Sensitivity Analysis in Meta-Analysis.
https://rdrr.io/cran/metasens/

\bibitem{key-1-1-1-1-13} Rufibach K. selectMeta: estimation of Weight
Functions in Meta Analysis. 2015. DOI: https://CRAN.R-project.org/package=selectMeta.

\bibitem{key-1-1-1-1-14} Friedrich J.O., Adhikari N.K.J. \& Beyene
J. The ratio of means method as an alternative to mean differences
for analyzing continuous outcome variables in meta-analysis: A simulation
study. \textit{BMC Med. Res. Methodol.} 2008, 8:32. DOI:10.1186/1471-2288-8-32.

\bibitem{key-1-2-1-1} Ackerman J.M., Nocera C.C., Bargh J.A. Incidental
haptic sensations influence social judgments and decisions. Science
2010, (80-: ) 328: 1712\textendash 1715.

\bibitem{key-1-2-1-2} Rabelo A.L.A., Keller V.N., Pilati R., Wicherts
J.M. No Effect of Weight on Judgments of Importance in the Moral Domain
and Evidence of Publication Bias from a Meta- Analysis. PLoS ONE 2015,
10(8): e0134808.

\bibitem{key-1-2-1-3} Francis G., Tanzman J., Matthews W.J. Excess
success for psychology articles in the journal Science. PLoS One 2014,
9: e114255.

\bibitem{key-1-1-1-1-15} Ning J., Chen Y. and Piao J. Maximum likelihood
estimation and EM algorithm of copas-like selection model for publication
bias correction. \textit{Biostatistics} 2017; 18(3):495\textendash 504.
DOI:10.1093/biostatistics/kxx004.

\bibitem{key-1-1-1-1-16} Moscati R., Jehle D., Ellis D., Fiorello
A., Landi M. Positive-outcome bias: comparison of emergency medicine
and general medicine literatures. \textit{Acad Emerg Med}. 1994;1(3):267\textendash 271.

\bibitem{key-1-1-1-1-17} Satterthwaite F.E. An approximate distribution
of estimates of variance components. \textit{Biometrics Bulletin}
1946;2(6):110-114.

\end{thebibliography}
\end{document}